\documentclass[10pt,conference,letterpaper]{IEEEtran}
\IEEEoverridecommandlockouts
\usepackage{cite}
\usepackage{amsmath,amssymb,amsfonts}
\usepackage{algorithmic}
\usepackage{graphicx}
\usepackage{textcomp}
\usepackage{listings}
\usepackage{xcolor}
\usepackage{balance}
\usepackage{subcaption}
\def\BibTeX{{\rm B\kern-.05em{\sc i\kern-.025em b}\kern-.08em
    T\kern-.1667em\lower.7ex\hbox{E}\kern-.125emX}}
\DeclareUnicodeCharacter{02BB}{{okina}}
\begin{document}

\title{CloudSec: An Extensible Automated Reasoning Framework for Cloud Security Policies

\thanks{This material is based upon work supported by the National Science Foundation Office of Advanced CyberInfrastructure, Collaborative Proposal: Frameworks: Project Tapis: Next Generation Software for Distributed Research(Award \#1931439)}
}
\makeatletter
\newcommand{\linebreakand}{%
  \end{@IEEEauthorhalign}
  \hfill\mbox{}\par
  \mbox{}\hfill\begin{@IEEEauthorhalign}
}
\makeatother

\author{\IEEEauthorblockN{Joe Stubbs*}
\IEEEauthorblockA{\textit{Texas Advanced Computing Center} \\
\textit{University of Texas at Austin}\\
Austin, TX, USA \\
jstubbs@tacc.utexas.edu}
\and
\IEEEauthorblockN{Smruti Padhy*}
\IEEEauthorblockA{\textit{Texas Advanced Computing Center} \\
\textit{University of Texas at Austin}\\
Austin, TX, USA \\
spadhy@tacc.utexas.edu}
\and
\IEEEauthorblockN{Richard Cardone}
\IEEEauthorblockA{\textit{Texas Advanced Computing Center} \\
\textit{University of Texas at Austin}\\
Austin, TX, USA \\
rcardone@tacc.utexas.edu}
\linebreakand %

\IEEEauthorblockN{Steven Black}
\IEEEauthorblockA{\textit{Texas Advanced Computing Center} \\
\textit{University of Texas at Austin}\\
Austin, TX, USA \\
scblack@tacc.utexas.edu}
\thanks{*The first two authors contributed equally to this work.}
}


\maketitle

\begin{abstract}
Users increasingly create, manage and share digital resources, including sensitive data, via cloud platforms and APIs. Platforms encode the rules governing access to these resources, referred to as \textit{security policies}, using different systems and semantics. As the number of resources and rules grows, the challenge of reasoning about them collectively increases. Formal methods tools, such as Satisfiability Modulo Theories (SMT) libraries, can be used to automate the analysis of security policies, but several challenges, including the highly specialized, technical nature of the libraries as well as their variable performance, prevent their broad adoption in cloud systems. In this paper, we present CloudSec, an extensible framework for reasoning about cloud security policies using SMT. CloudSec provides a high-level API that can be used to encode different types of cloud security policies without knowledge of SMT. Further, it is trivial for applications written with CloudSec to utilize and switch between different SMT libraries such as Z3 and CVC5. We demonstrate the use of CloudSec to analyze security policies in Tapis, a cloud-based API for distributed computational research used by tens of thousands of researchers.
\end{abstract}


\section{Introduction}
Through the use of cloud-based applications and services, users create valuable digital assets that must be secured. Each cloud platform makes use of a system for managing and enforcing the rules regarding which users have access to which digital assets, and there is little standardization across the vast number of such systems. For example, each of the major cloud computing providers have their own, independent systems for access management: Amazon Web Services makes use of AWS IAM \cite{AwsIam}, Google Cloud Platform uses Google IAM \cite{GoogleIAM}, and Microsoft Azure provides Azure Role Based Access Control (RBAC) \cite{AzureRBAC}. Kubernetes, the popular container orchestration system, has its own RBAC policy system \cite{K8sRBAC}. There are also many popular open source projects providing these capabilities, including Casbin \cite{Casbin}, KeyCloak \cite{KeyCloak}, Open Policy Agent \cite{OpenPolicyAgent}, etc. 

As systems evolve over time and as the number of digital assets and rules, referred to as \textit{security policies}, grows, ensuring policies are correctly written and match what is being enforced becomes more difficult. In particular, when the relationship between a user and a system changes -- for example, because the user is taking on new responsibilities as part of their job or changing jobs altogether -- entire sets of security rules often must be updated, and it can be very difficult to be sure the updated set enforces the correct access controls. 

Software based on formal methods such as Satisfiability Modulo Theories (SMT) ( \cite{cocBook},\cite{kroeninBook}) libraries provide techniques for automatically reasoning about entire collections of policies, and to prove or find counter-examples to mathematically precise statements regarding the policies. For example, in a cloud system that provides access to different types of files stored across various backends, SMT can be used to determine if a set of security policies satisfies ``rules" such as: developers have access to code repositories but not other files, data scientists have access to datasets for model training but not code repositories, and project managers have access to testing results but not training data or code. However, the broad adoption of SMT in cloud systems faces the following challenges:
1) the use of these tools requires a sophisticated understanding of the underlying SMT; 2) a significant effort must be made to encode a particular security policy’s semantics into an SMT library; and 3) performance on different security policy sets varies across different SMT libraries and even different versions of the same library. 

To address these challenges, we built CloudSec \cite{cloudsec}, an extensible automated reasoning framework for cloud security policies that do not require any SMT knowledge to use or extend to new cloud policy systems. CloudSec defines a set of core \texttt{Components} -- basic data types for defining the semantics of a security policy -- as well as the \texttt{PolicyType} and \texttt{Policy} abstract classes. These are linked to SMT solvers through CloudSec's library of \texttt{backends}, which implement encodings of the generic \texttt{Components} data types as well as proof methods based on the functionality provided by the solver. The initial release of CloudSec \cite{cloudsec} includes support for the Z3 (\cite{mouraTacas2008}, \cite{z3url}) and CVC5 (\cite{cvc5url}, \cite{barbosaTacas2022}) libraries as \texttt{backends}. Using CloudSec's building blocks, the policy types for a new system can be defined in just a few lines of Python code without any understanding of SMT. 

As open-source software, CloudSec can be used to analyze the performance of different backend solvers transparently. 
In this paper, we present CloudSec API demonstrating the use of Z3 and CVC5 solvers as backends. Furthermore, we utilize CloudSec to analyze security policies within the Tapis API platform \cite{tapisRefJstubbs2021}, a modern cloud platform enabling automated, secure, collaborative research computing. Used by thousands of researchers to automate data management and analysis tasks, Tapis software components run across several institutions and support dozens of independent projects, or ``tenants" and utilizes a robust permissions model that includes hierarchical roles, permission “types” with schemas for many services, and a multi-tenant authentication system. This multi-site, multi-tenant architecture compounds the challenge of ensuring that resources are not accidentally exposed to unauthorized parties. Using CloudSec we automatically establish or find counter-examples to policy requirements across entire sets of thousands of Tapis permissions.

In summary, the main contributions of this paper are:
\begin{itemize}
    \item[1.] Design and implementation of CloudSec, an extensible Python framework for leveraging SMT for security policy analysis with a user-friendly interface.
    \item[3.] Description of CloudSec usage in Tapis, a real-world cloud platform used by thousands of researchers.
\end{itemize}

The rest of the paper is organized as follows:
In Section 2, we provide background material on Tapis; in Section 3, we describe the CloudSec design, its use in Tapis, and give examples of policy types and encodings; in Section 4, we briefly discuss the initial performance results of CloudSec; in Section 5 we discuss the related work; and we conclude in Section 6 and outline some areas for future work.


\section{Background}
In this section, we provide background information on topics used throughout the rest of this paper. 

\subsection{Tapis}
Tapis \cite{tapisRefJstubbs2021} is a web-friendly, application programming interface (API) for research computing, allowing users to automate their interactions with advanced storage and computing resources in cloud and HPC datacenters. 
Primary Tapis features include a full-featured data management service, with synchronous endpoints for data ingest and retrieval as well as a reliable asynchronous data transfer facility; workload scheduling and code execution; a highly-scalable document store and metadata API; and support for streaming IoT/sensor data. 
Tapis supports reproducibility by recording a detailed data provenance and computation history of actions taken in the platform. Additionally, Tapis enables collaboration via a fine-grained permissions model, allowing data, metadata and computations to be kept private, shared with specific individuals or disseminated to entire research communities. Tapis has been used by thousands of researchers across projects funded by a number of government agencies, including CDC, DARPA, NASA, NIH, and NSF. 

\subsection{Tapis Security Policies}
Tapis is organized as a set of 14 independent HTTP web services (sometimes called \textit{microservices}) that coordinate together to accomplish larger tasks. The Tapis Security Kernel (SK) manages all authorization data -- information specifying which users have access to which Tapis objects and at what access level. SK stores this authorization data as \textit{permission} objects and provides HTTP endpoints for creating, retrieving, and modifying permissions. 


\section{Approach}
As mentioned in the Introduction, CloudSec provides a toolkit in the form of a Python library for utilizing SMT solvers to analyze security policies in real-world systems such as Tapis. The primary goal of CloudSec is to reduce the expertise needed to apply SMT technology to the study of security policies.

\begin{figure}%
    \centering
    {{\includegraphics[width=10cm]{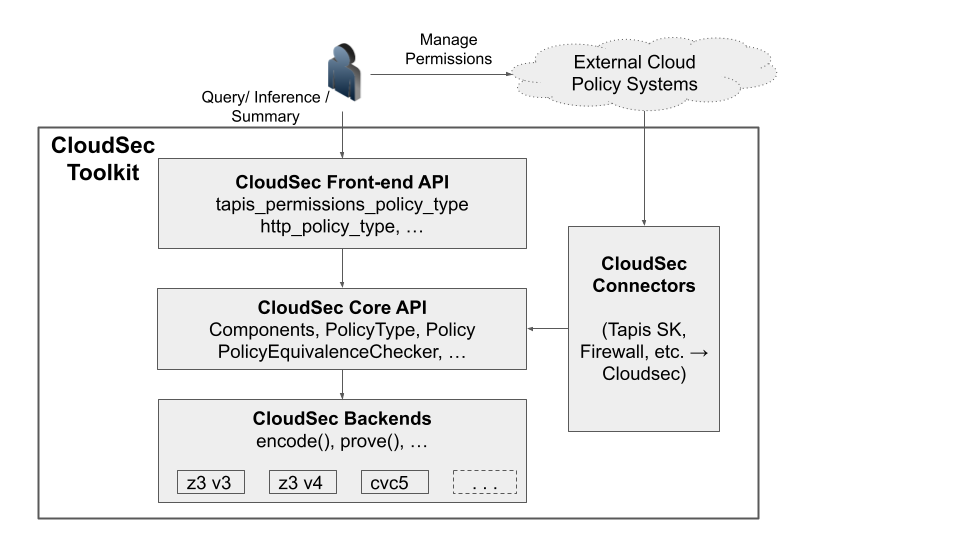}}}%
    \caption{CloudSec Overview and Usage Model}%
    \label{fig:cloudsec-overview}%
\end{figure}

\subsection{Design Challenge}


The primary design challenge faced by CloudSec was to provide abstractions with sufficient expressive power to support a large number of policy types while not requiring knowledge of SMT. CloudSec achieves this goal by decoupling the interface used for policy encoding from that used for policy analysis.

\subsection{CloudSec - Extensible Framework Design}
The primary components of the CloudSec framework include (i) the \texttt{core} module, with basic data types that can be used to build encodings of real-world security policy systems; (ii) the \texttt{backends} library, which provides implementations of encodings  of the types provided in the \texttt{core} module as well as proof methods for analyzing sets of policies via various solvers, such as Z3 and cvc5; (iii) the \texttt{cloud} module, which utilizes the types provided in \texttt{core} to define ready-to-use policy types for real systems; and (iv) the \texttt{connectors} library, which provides functions for converting security rules in source systems to CloudSec policies.  

Central to the design of CloudSec are the concepts of \texttt{Component}, \texttt{PolicyType}, and \texttt{Policy}, provided by the \texttt{core} module. Each \texttt{Component} type contains a data type, such as a \texttt{String},  \texttt{Enumeration}, \texttt{IP Address} or \texttt{Tuple}, the set of allowable values for the data type, and a matching strategy, which defines a ``match" relation on two values from the set of allowable values for the data type. Examples of currently supported matching strategies include exact matching and wildcard matching. \texttt{PolicyType}s build on the notion of \texttt{Component}s, with each \texttt{PolicyType} defining a list of \texttt{Component}s that it is comprised of. Finally, a \texttt{Policy} object represents a specific value for a given \texttt{PolicyType}. 

Using these primary notions, one can create \texttt{PolicyType}s comprised of \texttt{Components} for real-world systems in just a few lines of Python without requiring any knowledge of SMT. The \texttt{cloud} module provides examples of \texttt{PolicyType} objects, including a \texttt{tapis\_files\_policy\_type}, used to represent policies related to file objects in Tapis, and an \texttt{http\_api\_policy\_type}, which can be used to model policies in an arbitrary HTTP API. Moreover, by decoupling the definitions of \texttt{Components} and \texttt{PolicyTypes} from their implementations in different backend solvers, CloudSec provides a highly-extensible system in which support for additional solvers can be added independently of defining policy types for new systems.  

\subsection{An Example Policy Type and Policy Definition}
As a first example, we describe the \texttt{http\_api\_policy\_type}, available in the CloudSec toolkit. We created this policy type to illustrate what can be achieved with CloudSec. The policy type represents policies governing access to resources defined in a multi-tenant, microservice API platform. The policy type is comprised of three components:
\textit{principal}, \textit{resource}, and \textit{action}. A principal is a \texttt{Tuple} component type with two fields, \texttt{tenant} and \texttt{username}, representing a unique user identity in the platform to whom access to some resource is being allowed or denied by the policy. A tenant is a \texttt{StringEnum} component with enumerated values representing all possible tenants. We chose a \texttt{StringEnum} to model the tenant under the assumption that the number of tenants would be relatively small. On the other hand, the username field is modeled with a \texttt{String} component type defined over an alphanumeric set and maximum length.

Similarly, a resource is modeled using a \texttt{Tuple} component type with three fields, \textit{tenant, service}, and \textit{path}, that collectively distinguish a specific HTTP endpoint (i.e., HTTP resource) within a service to which access is being allowed or denied in some tenant. A service is a \texttt{StringEnum} component with enumerated values representing the names of the APIs in the platform. The path is a String component defined over a path character set with a maximum length. The path represents the URL path corresponding to the resource. Finally, the action is a \texttt{StringEnum} that denotes the HTTP method being authorized or not authorized, such as \{``GET'',``POST'',``PUT'',``DELETE''\}. 
An example policy definition of \texttt{http\_api\_policy\_type} is shown below:
\begin{lstlisting}[language=Python]
p = Policy(policy_type=http_api_policy_type, 
    principal=(``a2cps",``jdoe"), 
    resource=(``a2cps",``files", 
    ``/ls6/home/jdoe"),action=``GET",
    decision=``allow")
\end{lstlisting}

Note that all policy types have a special \texttt{decision} field which is a \texttt{StringEnum} with values ``allow" and ``deny". 

\subsection{Tapis Policy Types}
The CloudSec toolkit includes policy types representing permissions objects in the Tapis API platform. For example, we provide the \texttt{tapis\_files\_policy\_type} for dealing with Tapis permissions related to file objects. The \texttt{tapis\_file\_perm} component is a \texttt{Tuple} type representing the Tapis files ``perm spec" (see \cite{TapisPerms}) and includes fields for the tenant, system id, permission level, and file path. Because of the simple and flexible CloudSec core API, the entirety of the Tapis policy type implementations constitutes less than 20 lines of Python code.

\subsection{Translating a Policy Set to SMT Formula}
\label{subsec:smt_encoding}
A policy set is defined as a set of policies, i.e., $PS = \{ P_1, P_2, \cdot\cdot\cdot, P_i, \cdot\cdot\cdot, P_{n-1},P_n\}$ where $1\leq i \leq n $. Each policy, $P_i = (c_1, c_2, \cdot\cdot\cdot, c_j, \cdot\cdot\cdot, c_{m-1},c_m, Decision)$ where $1 \leq j \leq m$. $c_j$ denotes the value for \texttt{Component} $C_j$ of Policy $P_i$. Note that a policy type determines the number and type of \texttt{Components} in a policy. $Decision$ is a value from the set $\{``allow", ``deny"\}$ to denote if a policy allows or denies access. 
A policy is translated to an SMT formula as:
\begin{equation*}
    \mathcal{P} = \bigwedge_{j=1}^{m}\left (\bigvee_{c\in C_j(P)}C_j=c \right )
\end{equation*}
$C_j(P)$ denotes set of values defined for \texttt{Component} $C_j$ in a policy $P$. If $C_j(P)$ is a \texttt{Tuple} component type with $k$ components , let say, $(t_1, t_2, ..,t_k) $, then it is further encoded as $\bigwedge_{l=1}^{k}\left (t_k =v \right ) $ where component $t_k$ takes one of the component allowed values, $v$.

A policy set can be SMT encoded as: 
\begin{equation*}
    \mathcal{PS} =  \left(\bigvee_{AllowSet}\mathcal{P}\right) \bigwedge \neg\left(\bigvee_{DenySet}\mathcal{P}\right)
\end{equation*}
Where $AllowSet = \{\forall P \in PS: P.Decision = ``allow"\}$ and $DenySet = \{\forall P \in PS: P.Decision = ``deny"\} $

\subsection{Connectors and Converting SK Policies to Cloudsec}
\label{subsec:policy_converter}
 using CloudSec to analyze security rules from a real-world system requires one to generate CloudSec policy objects (of the appropriate type) from authorization data residing in the external system. A CloudSec \texttt{connector} can be written for the external system to simplify this effort. The CloudSec toolkit currently includes a connector for Tapis files permissions, allowing CloudSec \texttt{tapis\_files\_policy\_type} policies to be generated from Tapis permissions data with a single function call. 

Using CloudSec, we developed a program that generates Tapis files policy objects from all Tapis files permissions in the SK for a configurable set of users. The program then uses CloudSec solver backends to prove that the source policies conform to certain rules or find counter examples. For instance, using our program, we analyzed policies for users within a certain project built with Tapis, generating over 3,000 CloudSec policies. We then were able to prove that for this project, no users except for the users in the admin-scientist role had read/write access to files in a protected \texttt{data} directory on a specific Tapis system. Similarly, we prove that public access to generated result files is restricted to files with a \texttt{.png} extension. 

Establishing these kinds of results harnesses the full power of CloudSec -- while the Tapis SK is highly efficient at answering the question, ``does a given user have access to a specific resource?", it cannot reason about an entire set of permissions at once. Furthermore, our program must use both Z3 and cvc5 to find proofs like the ones above, as some proofs could not be found by one solver or the other. 

We did an initial performance evaluation of the CloudSec toolkit 
1) to establish that the CloudSec software was a viable approach (i.e., can be done on commodity hardware in a reasonable amount of time) to analyze policies of the sizes that show up in real systems such as Tapis (i.e., on the order of a few thousand policies) (Section \ref{subsec:policy_converter}); and 2) to show that CloudSec could be used as a framework for comparing different SMT backends for specific analyses \cite{cloudsec}. 
We also observed that there are scenarios for which either Z3 or CVC5 or both, along with their different versions exhibit performance cliffs. 

\section{Performance}
We did a performance evaluation of the CloudSec toolkit 
1) to establish that the CloudSec software was a viable approach (i.e., can be done on commodity hardware in a reasonable amount of time) to analyze policies of the sizes that show up in real systems such as Tapis (i.e., on the order of a few thousand policies) (Section \ref{subsec:policy_converter}); and 2) to show that CloudSec could be used as a framework for comparing different SMT backends for specific analyses \cite{cloudsec}. 
We also observed that there are scenarios for which either Z3 or CVC5 or both, along with their different versions exhibit performance cliffs. We provide the details of the performance evaluation in subsequent subsections.

We measured the performance to check trivial implications for two policy sets as the number of policies grows. We defined different components (\texttt{String} and \texttt{StringEnum}) and policy types using those components to create policy sets. Each policy set consists of policies of the same policy type. We repeated the test for both Z3 and cvc5 backends. We ran the tests on a machine with the following configuration: 32 CPUs, Intel(R) Xeon(R) CPU E5-2660 0 @ 2.20GHz, 128 GB RAM.

\subsection{StringEnum Scalability}
We defined a policy type containing a \texttt{StringEnum} with a variable number, $N$, allowable values and with wildcard matching. For example, with $N=5$, the policy type's \texttt{StringEnum} will have allowable values \{ `0', `1',`2',`3',`4'\}.   
We created two policy sets, $P$ and $Q$. $P$ contained a policy $P_i$ for every allowable value, i.e.,  $P = \cup_{i=0}^{N-1}\{(P_i, ``allow")\}$ where $P_i=i$. $Q$ contained a single wildcard policy, $Q=\{(*, ``allow")\}$. We varied the number of unique elements, $N$, in the range $10\leq N \leq 4000$. We measured the time to perform: 1) data load, 2) SMT encoding, 3) $P \implies Q$ and 4) $Q \implies P$ for Z3 as well as for cvc5. In Figure \ref{fig:string-enum} a, we observed that when using the Z3 backend, the SMT encoding, $P\implies Q$ and $Q\implies P$ took similar amounts of time, while the data load took significantly less time. When we used cvc5, SMT encoding took the most of the computation time while the data load time was more than the implication prove time. For this performance test, cvc5 was roughly 90\% faster than Z3 in total time.  

\begin{figure*}%
    \centering
    \subfloat[\centering Using Z3 backend]{{\includegraphics[width=5.6cm]
    {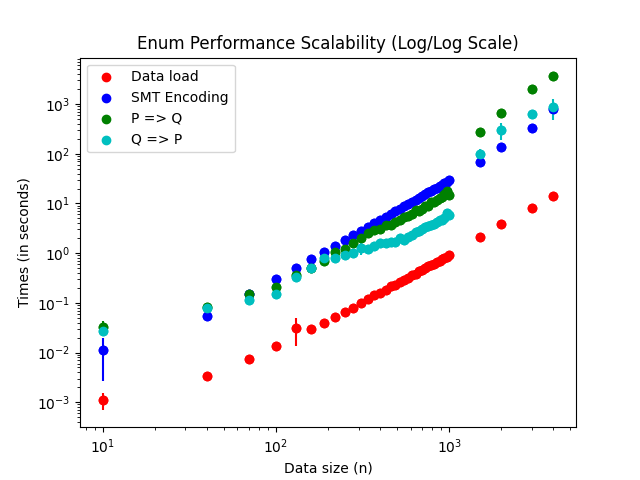} }}%
    \qquad
    \subfloat[\centering Using cvc5 backend]{{\includegraphics[width=5.6cm]
    {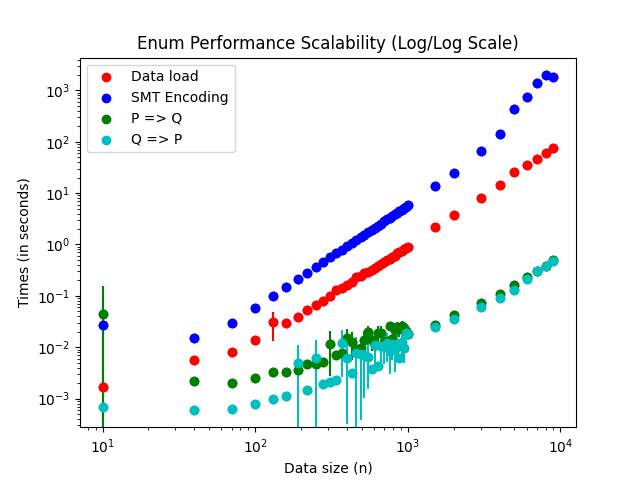} }}%
    \caption{Dynamic StringEnum Policy type Scalability}%
    \label{fig:string-enum}%
\end{figure*}

\subsection{String with wildcard Scalability}
We defined a policy type with a single \texttt{String} component with wildcard matching. The string component was defined over a character set that included alphanumeric characters, `/', and the wildcard character `*'. The maximum length of the string was 100. We created two policy sets, $P$ and $Q$, for each $10 \leq N \leq 1000$. The policy set $P =\cup_{i=0}^{N-1}\{(P_i, ``allow")\}$ where each $P_i$ was defined to be the static string $``a1b2c3d4e5/i"$. The policy set $Q=\cup_{i=0}^{N-1}\{(Q_i, ``allow")\}$ where each $Q_i$ is defined as the string $``a1b2c3d4e5/i*"$ ending in the wildcard character. For example, if $N=2$, $P = \{``a1b2c3d4e5/1", ``a1b2c3d4e5/2"\}$ and $Q = \{``a1b2c3d4e5/1*", ``a1b2c3d4e5/2*"\}$. In this case, there are 2 policies in $P$ and two policies in $Q$. 

We varied N and  measured the data load, SMT encoding, and $P \implies Q$ times for Z3 as well as for cvc5. Note $Q \implies P$ is not valid.
In Figure \ref{fig:string-wc}, we observe that Z3 was faster than cvc5 in total time and, in particular, in proving $P \implies Q$, Z3 was often an order of magnitude or more faster. Both backends follow the same pattern where data load and SMT encoding times are less than implication prove time. Z3 shows good performance even for $N=1000$, which is 1000 policies in $P$ and in $Q$.
\begin{figure*}%
    \centering
    \subfloat[\centering Using Z3 backend]{{\includegraphics[width=5cm]
    {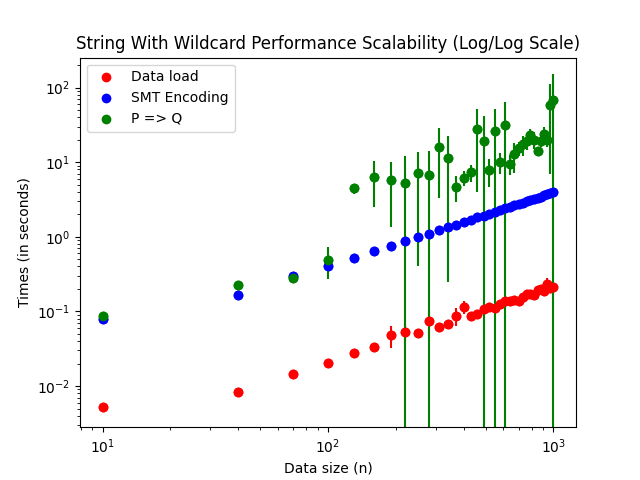 }}}%
    \qquad
    \subfloat[\centering Using cvc5 backend]{{\includegraphics[width=5cm]
    {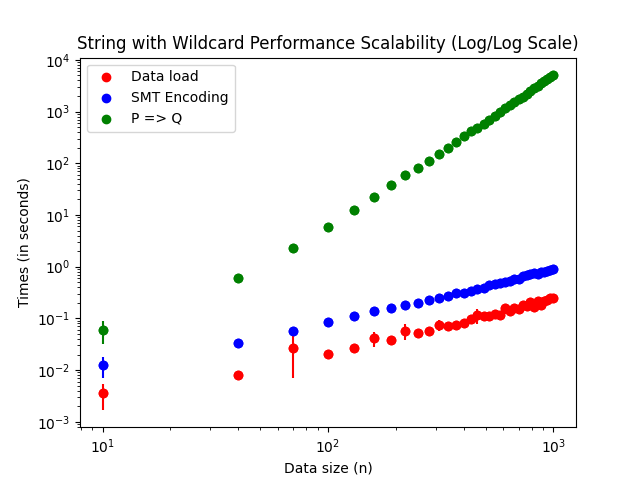}}}%
    \caption{String with Wildcard Policy type Scalability}%
    \label{fig:string-wc}%
\end{figure*}

\subsection{Performance Cliffs for SMT Solvers}
\label{subsec:PerfCliffs}
We observed some performance cliffs while using Z3 and cvc5. First, consider two simple policy sets in which the policy type has just one component, \textit{path}, of string type. Let $P={(``/sys1*", ``allow")}$ and $Q = {(``/*", ``allow")}$. While trying to prove $P$ is less permissive than $Q$, the Z3 backend hangs but cvc5 is able to prove it.

Second, consider the policy sets:
\begin{align*}
P=\{(``jstubbs",``s2/home/jstubbs/*", ``*", ``allow"), \\(``jstubbs",``s2/*",``PUT", ``deny"),\\
(``jstubbs",``s2/*",``POST", ``deny")\}
\end{align*}
 and
 \begin{align*}
Q=\{(``jstubbs",``s2/home/jstubbs/a.out",``GET", \\ ``allow"), \\(``jstubbs",``s2/home/jstubbs/b.out",``GET", ``allow")\}
\end{align*}
Z3 was able to to prove $Q \implies P$ but cvc5 hangs. 
\section{Related work}
There have been several works on cloud-access control policies analysis using SMT (\cite{backesFmcad2018}, \cite{LiuIcet2022}, \cite{rungtaCav2022}, \cite{eiersIcse2022}). Our work is closely related to the approach used in Zelkova (\cite{backesFmcad2018},  \cite{rungtaCav2022}). Zelkova is an AWS policy analysis tool that verifies AWS policies by reasoning if a policy is less or equally permissive than the other. It encodes AWS policies to SMT formulas and uses SMT solvers, Z3, CVC, and Z3AUTOMATA, to verify and prove the properties. Zelkova is not open-source and cannot be extended to other policy languages. CloudSec's approach is similar to Zelkova and other prior works in defining a policy language and translating policies to SMT formulas for reasoning using SMT solvers. The main differences are that 1) CloudSec is an open-source, extensible automated reasoning framework where users or developers can define their policy types not restricted to cloud policies depending on their application, and 2) developers can plug different SMT solvers into CloudSec's extensible backend. Additionally, by defining a policy converter, CloudSec can be easily integrated into existing cloud-hosted services, as demonstrated in section \ref{subsec:policy_converter}. 

Eiers et al. \cite{eiersIcse2022} proposed a framework to quantify the permissiveness of access policies using model counting constraint solvers and relative permissiveness between the policies. They also built an open-source tool, QUACKY, that analyzes AWS and Azure policies. The CloudSec framework focuses on easy extensibility, defining different policy types, and making the reasoning tool more accessible to users. We envision that QUACKY could potentially be a component in the CloudSec framework to compute relative permissiveness. 

Several works have studied the verification of network access control, connectivity, and configuration policies. Jayaraman et al. (\cite{jayaramanMicrosoft2014}, \cite{jayaramanSigcomm2019}) proposed and built a tool called SECGURU that automatically validates network connectivity policies using SMT bit vector theory and the Z3 solver. SECGURU is a closed source tool used in Azure.
 Fogel et al. \cite{fogelNsdi2015} proposed and developed an open-source tool, Batfish, that analyzes network configuration and detects errors. Campion \cite{tangSigcomm2021} is another open-source tool for debugging router configuration and has been implemented as an extension to Batfish. It localizes crucial errors to relevant configuration lines. Beckett et al. \cite{beckettSigcomm2017} proposed a general approach for network configuration that translates both control and data plane behaviors into a logical formula and use SMT solver, Z3, to verify the properties. They have implemented this approach in the tool called Minesweeper. CloudSec could be extended to support network access policies such as Firewall and router policies.

\section{Conclusion and Future Work}
In this paper, we presented a description of the CloudSec library, which simplifies the use of SMT in analyzing security policies in real-world systems. We applied CloudSec to the Tapis API platform to analyze thousands of permissions records at once. In the future, we plan to incorporate CloudSec into a public API within Tapis, allowing any user to easily submit security analysis jobs using HTTP requests. Further, we will explore adding support for additional backends to the project and applying CloudSec to additional real-world systems, such as AWS, and Kubernetes. 
\balance
\bibliographystyle{IEEEtran}
\bibliography{IEEEabrv,secfm}

\begin{thebibliography}{10}
\providecommand{\url}[1]{#1}
\csname url@samestyle\endcsname
\providecommand{\newblock}{\relax}
\providecommand{\bibinfo}[2]{#2}
\providecommand{\BIBentrySTDinterwordspacing}{\spaceskip=0pt\relax}
\providecommand{\BIBentryALTinterwordstretchfactor}{4}
\providecommand{\BIBentryALTinterwordspacing}{\spaceskip=\fontdimen2\font plus
\BIBentryALTinterwordstretchfactor\fontdimen3\font minus
  \fontdimen4\font\relax}
\providecommand{\BIBforeignlanguage}[2]{{%
\expandafter\ifx\csname l@#1\endcsname\relax
\typeout{** WARNING: IEEEtran.bst: No hyphenation pattern has been}%
\typeout{** loaded for the language `#1'. Using the pattern for}%
\typeout{** the default language instead.}%
\else
\language=\csname l@#1\endcsname
\fi
#2}}
\providecommand{\BIBdecl}{\relax}
\BIBdecl

\bibitem{AwsIam}
\BIBentryALTinterwordspacing
{AWS IAM}. (2022, Dec.) {AWS IAM}. [Online]. Available:
  \url{https://docs.aws.amazon.com/IAM/latest/UserGuide/access.html}
\BIBentrySTDinterwordspacing

\bibitem{GoogleIAM}
\BIBentryALTinterwordspacing
{Google IAM}. (2022, Dec.) {Google IAM}. [Online]. Available:
  \url{https://cloud.google.com/iam/}
\BIBentrySTDinterwordspacing

\bibitem{AzureRBAC}
\BIBentryALTinterwordspacing
AzureRBAC. (2022, Dec.) {AzureRBAC}. [Online]. Available:
  \url{https://learn.microsoft.com/en-us/azure/role-based-access-control/overview}
\BIBentrySTDinterwordspacing

\bibitem{K8sRBAC}
\BIBentryALTinterwordspacing
{Kubernetes}. (2023, May) {Kubernetes RBAC}. [Online]. Available:
  \url{https://kubernetes.io/docs/reference/access-authn-authz/rbac/}
\BIBentrySTDinterwordspacing

\bibitem{Casbin}
\BIBentryALTinterwordspacing
Casbin. (2022, Dec.) Casbin. [Online]. Available: \url{https://casbin.org/}
\BIBentrySTDinterwordspacing

\bibitem{KeyCloak}
\BIBentryALTinterwordspacing
{KeyCloak}. (2022, Dec.) {KeyCloak}. [Online]. Available:
  \url{https://www.keycloak.org/}
\BIBentrySTDinterwordspacing

\bibitem{OpenPolicyAgent}
\BIBentryALTinterwordspacing
{Open Policy Agent}. (2022, Dec.) {Open Policy Agent}. [Online]. Available:
  \url{https://www.openpolicyagent.org/}
\BIBentrySTDinterwordspacing

\bibitem{cocBook}
A.~R. Bradley and Z.~Manna, \emph{{The Calculus of Computation: Decision
  Procedures with Applications to Verification}}.\hskip 1em plus 0.5em minus
  0.4em\relax Springer, 2017.

\bibitem{kroeninBook}
D.~Kroening and O.~Strichman, \emph{{Decision Procedures: An Algorithmic Point
  of View}}, ser. Texts in Theoretical Computer Science. An EATCS Series.\hskip
  1em plus 0.5em minus 0.4em\relax Springer, 2017.

\bibitem{cloudsec}
\BIBentryALTinterwordspacing
``{CloudSec},'' (Date last accessed May-21-2023). [Online]. Available:
  \url{https://github.com/applyfmsec/cloudsec}
\BIBentrySTDinterwordspacing

\bibitem{mouraTacas2008}
L.~de~Moura and N.~Bj{\o}rner, ``Z3: An efficient smt solver,'' in \emph{Tools
  and Algorithms for the Construction and Analysis of Systems}, C.~R.
  Ramakrishnan and J.~Rehof, Eds.\hskip 1em plus 0.5em minus 0.4em\relax
  Berlin, Heidelberg: Springer Berlin Heidelberg, 2008, pp. 337--340.

\bibitem{z3url}
\BIBentryALTinterwordspacing
``Z3,'' (Date last accessed 4-December-2022). [Online]. Available:
  \url{https://github.com/Z3Prover/z3}
\BIBentrySTDinterwordspacing

\bibitem{cvc5url}
\BIBentryALTinterwordspacing
``{CVC5: An efficient open-source automatic theorem prover for satisfiability
  modulo theories (SMT) problems.}'' (Date last accessed 4-December-2022).
  [Online]. Available: \url{https://cvc5.github.io/}
\BIBentrySTDinterwordspacing

\bibitem{barbosaTacas2022}
H.~Barbosa, C.~W. Barrett, M.~Brain, G.~Kremer, H.~Lachnitt, M.~Mann,
  A.~Mohamed, M.~Mohamed, A.~Niemetz, A.~N{\"{o}}tzli, A.~Ozdemir, M.~Preiner,
  A.~Reynolds, Y.~Sheng, C.~Tinelli, and Y.~Zohar, ``cvc5: {A} versatile and
  industrial-strength {SMT} solver,'' in \emph{Tools and Algorithms for the
  Construction and Analysis of Systems - 28th International Conference, {TACAS}
  2022, Held as Part of the European Joint Conferences on Theory and Practice
  of Software, {ETAPS} 2022, Munich, Germany, April 2-7, 2022, Proceedings,
  Part {I}}, ser. Lecture Notes in Computer Science, D.~Fisman and G.~Rosu,
  Eds., vol. 13243.\hskip 1em plus 0.5em minus 0.4em\relax Springer, 2022, pp.
  415--442.

\bibitem{tapisRefJstubbs2021}
J.~Stubbs, R.~Cardone, M.~Packard, A.~Jamthe, S.~Padhy, S.~Terry, J.~Looney,
  J.~Meiring, S.~Black, M.~Dahan, S.~Cleveland, and G.~Jacobs, ``{Tapis: An API
  Platform for Reproducible, Distributed Computational Research},'' in
  \emph{Advances in Information and Communication FICC 2021}.\hskip 1em plus
  0.5em minus 0.4em\relax Springer International Publishing, 2021, pp.
  878--900.

\bibitem{TapisPerms}
\BIBentryALTinterwordspacing
Tapis. (2022, Dec.) Tapis permissions. [Online]. Available:
  \url{https://tapis.readthedocs.io/en/latest/technical/security.html\#id3}
\BIBentrySTDinterwordspacing

\bibitem{backesFmcad2018}
J.~Backes, P.~Bolignano, B.~Cook, C.~Dodge, A.~Gacek, K.~Luckow, N.~Rungta,
  O.~Tkachuk, and C.~Varming, ``Semantic-based automated reasoning for aws
  access policies using smt,'' in \emph{2018 Formal Methods in Computer Aided
  Design (FMCAD)}, 2018, pp. 1--9.

\bibitem{LiuIcet2022}
A.~Liu, X.~Du, N.~Wang, X.~Wang, X.~Wu, and J.~Zhou, ``Implement security
  analysis of access control policy based on constraint by smt,'' in \emph{2022
  IEEE 5th International Conference on Electronics Technology (ICET)}, 2022,
  pp. 1043--1049.

\bibitem{rungtaCav2022}
N.~Rungta, ``A billion smt queries a day (invited paper),'' in \emph{Computer
  Aided Verification}, S.~Shoham and Y.~Vizel, Eds.\hskip 1em plus 0.5em minus
  0.4em\relax Cham: Springer International Publishing, 2022, pp. 3--18.

\bibitem{eiersIcse2022}
\BIBentryALTinterwordspacing
W.~Eiers, G.~Sankaran, A.~Li, E.~O'Mahony, B.~Prince, and T.~Bultan,
  ``Quantifying permissiveness of access control policies,'' in
  \emph{Proceedings of the 44th International Conference on Software
  Engineering}, ser. ICSE '22.\hskip 1em plus 0.5em minus 0.4em\relax New York,
  NY, USA: Association for Computing Machinery, 2022, p. 1805–1817. [Online].
  Available: \url{https://doi.org/10.1145/3510003.3510233}
\BIBentrySTDinterwordspacing

\bibitem{jayaramanMicrosoft2014}
K.~Jayaraman, N.~Bjørner, G.~Outhred, and C.~Kaufman, ``Automated analysis and
  debugging of network connectivity policies,'' Microsoft, Tech. Rep.
  MSR-TR-2014-102, July 2014.

\bibitem{jayaramanSigcomm2019}
\BIBentryALTinterwordspacing
K.~Jayaraman, N.~Bj\o{}rner, J.~Padhye, A.~Agrawal, A.~Bhargava, P.-A.~C.
  Bissonnette, S.~Foster, A.~Helwer, M.~Kasten, I.~Lee, A.~Namdhari, H.~Niaz,
  A.~Parkhi, H.~Pinnamraju, A.~Power, N.~M. Raje, and P.~Sharma, ``Validating
  datacenters at scale,'' in \emph{Proceedings of the ACM Special Interest
  Group on Data Communication}, ser. SIGCOMM '19.\hskip 1em plus 0.5em minus
  0.4em\relax New York, NY, USA: Association for Computing Machinery, 2019, p.
  200–213. [Online]. Available: \url{https://doi.org/10.1145/3341302.3342094}
\BIBentrySTDinterwordspacing

\bibitem{fogelNsdi2015}
A.~Fogel, S.~Fung, L.~Pedrosa, M.~Walraed-Sullivan, R.~Govindan, R.~Mahajan,
  and T.~Millstein, ``A general approach to network configuration analysis,''
  in \emph{Proceedings of the 12th USENIX Conference on Networked Systems
  Design and Implementation}, ser. NSDI'15.\hskip 1em plus 0.5em minus
  0.4em\relax USA: USENIX Association, 2015, p. 469–483.

\bibitem{tangSigcomm2021}
\BIBentryALTinterwordspacing
A.~Tang, S.~K.~R. Kakarla, R.~Beckett, E.~Zhai, M.~Brown, T.~Millstein,
  Y.~Tamir, and G.~Varghese, ``Campion: Debugging router configuration
  differences,'' in \emph{Proceedings of the 2021 ACM SIGCOMM 2021 Conference},
  ser. SIGCOMM '21.\hskip 1em plus 0.5em minus 0.4em\relax New York, NY, USA:
  Association for Computing Machinery, 2021, p. 748–761. [Online]. Available:
  \url{https://doi.org/10.1145/3452296.3472925}
\BIBentrySTDinterwordspacing

\bibitem{beckettSigcomm2017}
\BIBentryALTinterwordspacing
R.~Beckett, A.~Gupta, R.~Mahajan, and D.~Walker, ``A general approach to
  network configuration verification,'' in \emph{Proceedings of the Conference
  of the ACM Special Interest Group on Data Communication}, ser. SIGCOMM
  '17.\hskip 1em plus 0.5em minus 0.4em\relax New York, NY, USA: Association
  for Computing Machinery, 2017, p. 155–168. [Online]. Available:
  \url{https://doi.org/10.1145/3098822.3098834}
\BIBentrySTDinterwordspacing

\end{thebibliography}

\end{document}